# Enhancing synchronizability of weighted dynamical networks using betweenness centrality

**Mahdi Jalili, Ali Ajdari Rad, and Martin Hasler**

Ecole Polytechnique Fédérale de Lausanne, School of Computer and Communication Sciences, Laboratory of Nonlinear Systems, CH 1015, Lausanne, Switzerland

**ABSTRACT**

By considering the eigenratio of the Laplacian of the connection graph as synchronizability measure, we propose a procedure for weighting dynamical networks to enhance their synchronizability. The method is based on node and edge betweenness centrality measures and is tested on artificially constructed scale-free, Watts-Strogatz and random networks as well as on some real-world graphs. It is also numerically shown that the same procedure could be used to enhance the phase synchronizability of networks of nonidentical oscillators.

## I. INTRODUCTION

Complex networks are ubiquitous and networks of dynamical units serve as natural models for many real-world systems, with many examples ranging from Internet to the epidemiology, ecology, cell biology and social interactions [1, 2]. The interplay between structural properties of such complex networks on the one hand and dynamics located on their nodes on the other hand has attracted a great deal of attention [3-5]. Much of this interest is motivated by the fact that many real-world networks share some common structural properties such as small-world [6] and/or scale-free [7] attributes.

In recent years an avalanche of studies on the most conspicuous form of collective behavior, synchronization, has appeared [8]. Fundamental assumption of the most of the works in this field is that the individual dynamical systems are diffusively coupled with uniform strength over unweighted networks, but most relevant dynamical networks are inherently weighted and directed such as brain networks [9], ecological systems [10], traffic load of a road [11], social networks [12], metabolic networks [13] and technological networks [1, 14]. Thus, a natural question arises, namely: "Given a network, how one can assign proper connection weights to enhance its synchronizability?" This may provide us with insights into the behavior of real-world complex dynamical networks and guide us in designing large artificial networks. In technological networks with desirable synchronizability, assigning the appropriate interaction weights between dynamical units is important [14]. It has been recently shown that networks with properly assigned weights can be distinctly more synchronizable than unweighted networks [15].

In this paper, we give an algorithm for assigning connection weights to enhance the synchronizability of dynamical networks. Starting with a connected undirected and unweighted network, and by considering its local and global structural properties such as degree, node and edge betweenness centralities, we end up with an asymmetric weighted network with enhanced synchronizability. Compared to the other methods in this context [16-20], we give evidence that the proposed method leads to higher synchronizability for a class of scale-free, Watts-Strogatz and random networks as well as for many real-world complex networks. We also show that the same weighting procedure enhances the phase synchronizability of coupled nonidentical dynamical systems.

## II. WEIGHTING ALGORITHM

Let us start with a dynamical network of $N$ linearly coupled identical systems with the following equations of the motion

$$\dot{\mathbf{x}}_i = \mathbf{F}(\mathbf{x}_i) - \sigma \sum_{j=1}^{N} g_{ij} \mathbf{H}(\mathbf{x}_j) \ , \ i = 1, \ldots, N, \tag{1}$$

where $\mathbf{x}_i \in \mathbb{R}^d$ are the state vectors, $\mathbf{F}: \mathbb{R}^d \to \mathbb{R}^d$ defines the individual system's dynamical equation and $\sigma$ is the uniform coupling strength. The dynamical systems are coupled via a linear output function $\mathbf{H}: \mathbb{R}^d \to \mathbb{R}^d$ and the coupling matrix $G = (g_{ij})$. We assume that $G$ is symmetric, has non-positive off-diagonal elements and has zero row sums. It is the Laplacian matrix of the coupling graph.

The variational equations of manifold synchronized solution $(\mathbf{x}_i(t) = \mathbf{s}(t), \ \forall i)$ can be diagonalized into $N$ blocks of the form $\dot{\boldsymbol{\zeta}}_i = D\mathbf{F}(s(t))\boldsymbol{\zeta}_i - \sigma \lambda_i \mathbf{H}(\boldsymbol{\zeta}_i)$, where $\lambda_i$ are the eigenvalues of $G$, ordered as $0 = \lambda_1 \leq \lambda_2 \leq \ldots \leq \lambda_N$ and $\lambda_1$ is associated with the synchronized manifold. In the sequel, we suppose that the connection graph is connected, which implies the strict inequality $0 = \lambda_1 < \lambda_2$.

The largest Lyapunov exponent of the above variational equation, $\Lambda(\sigma \lambda_i)$, called master-stability-function [21], gives a sufficient condition for the local stability of the synchronization manifold: if the synchronization manifold is locally stable we must have $\Lambda(\sigma \lambda_i) < 0$, $i = 2, \ldots, N$. For a number of systems such as $x$–coupled Rössler oscillators, the master-stability-function is negative only within a bounded interval $(v_1, v_2)$ [21]. Requiring all coupling strengths lie within such an interval, i.e. $v_1 < \sigma \lambda_2 \leq \ldots \leq \sigma \lambda_N < v_2$, leads to the following condition for the local stability of the synchronization manifold: $\lambda_N / \lambda_2 < v_2 / v_1$. The left-hand side of the inequality depends solely on the structure of the graph, while the right-hand side depends on the dynamics of the individual systems and on the coupling configuration. There is a number of interpretations for synchronizability of dynamical networks [15]. We adopt here the following interpretation: the larger the range of parameters of the individual dynamical systems that allows for synchronization, the better the synchronizability of the network. This relates the synchronizability to the eigenratio $\lambda_N/\lambda_2$, and concludes that the smaller the eigenratio $\lambda_N/\lambda_2$ of a network, the better its synchronizability [22].

In general, there are two possible ways to enhance the synchronizability of dynamical networks: rewiring of the links [23] and/or assigning proper weights for the existing links [15]. For many applications it is not possible to change the network topology and the only option to enhance the synchronizability is weighting the links. In Networks with good synchronizability, couplings between the nodes are neither necessarily uniform nor symmetric. A very first attempt to assign the proper connection weights for enhancing the synchronizability was proposed in [18, 19], where the coupling on the right hand side of the equation (1) was taken as $\left(\sigma/k_i^\beta\right)\sum_{j\in N_i}\mathbf{H}(\mathbf{x}_i-\mathbf{x}_j)$, with $k_i$ the degree of node $i$ and $N_i$ the set of neighbors of node $i$. The value of $\beta = 1$ was found to be optimal in [19] for synchronizability. Further enhancement of the synchronizability was achieved by scaling the weight of each edge by its load [17], where the coupling takes the form as $\left(\sigma/\sum_{j\in N_i}\rho_{ij}^\beta\right)\sum_{j\in N_i}\rho_{ij}^\beta\mathbf{H}(\mathbf{x}_i-\mathbf{x}_j)$, where $\rho_{ij}$ is the load of the edge $e_{ij}$ between the $i$-th and $j$-th node. The load (also known as edge betweenness centrality) of $e_{ij}$ is defined as $\rho_{ij}=\sum_{p\neq q}\left(\Gamma_{pq}(e_{ij})/\Gamma_{pq}\right)$, where $\Gamma_{pq}$ is the number of shortest paths from the $p$-th to the $q$-th node and $\Gamma_{pq}(e_{ij})$ is the number of these paths making use of $e_{ij}$. The optimal condition $\beta = 1$ was found for synchronizability [17] that itself performs better than the optimal case of [19]. In this way, not only the local structural information but also the effects of network structure at a global level are taken into account. Very recently, by considering the concepts of gradient networks, another weighting algorithm as $\left(\sigma/\sum_{j\in N_i}k_j^\beta\right)\sum_{j\in N_i}k_j^\beta\mathbf{H}(\mathbf{x}_i-\mathbf{x}_j)$ has been proposed [20].

Here we show that further enhancement of synchronizability can be achieved by considering not only connection loads but also node betweenness centralities. Node betweenness centrality $C_i$ is a centrality measure of the $i$-th node in a graph, which counts the number of shortest paths making use of that node (except shortest paths between the $i$-th node and other nodes) [24]. More precisely, $C_i = \sum_{p \neq i \neq q} \left( \Gamma_{pq}(i) / \Gamma_{pq} \right)$, where $\Gamma_{pq}$ is the number of shortest paths from the $p$-th to the $q$-th node and $\Gamma_{pq}(i)$ is the number of these shortest paths making use of the $i$-th node. In the weighting procedure we propose, the weight of each edge will be a function of its load and the betweenness centrality of the tail node. The resulting weighted network becomes directed; the links go from the head nodes to the tail nodes. The weight of an edge should essentially be proportional to the betweenness centrality of the edge as proposed in [17]. The dependence of the weight to the betweenness centrality of the tail node is also straight forward. Nodes with high values of betweenness centrality can be regarded as hub nodes, i.e. many shortest paths make use of them. Therefore, it is reasonable to increase the weight of the links ending to these nodes. More precisely, the network equations read

$$\dot{\mathbf{x}}_i = \mathbf{F}(\mathbf{x}_i) - \frac{\sigma}{\sum_{j \in N_i} \left( \varepsilon + C_j^\alpha \right) \rho_{ij}} \sum_{j \in N_i} \left( \varepsilon + C_j^\alpha \right) \rho_{ij} \mathbf{H}(\mathbf{x}_i - \mathbf{x}_j) \quad ; \quad i = 1, 2, \ldots, N, \quad (2)$$

where $\alpha$ is a real tunable parameter, and $\varepsilon$ is a small positive value to make $(\varepsilon + C_j^\alpha) > 0$ (some nodes may have betweenness centrality equal to zero). Here, we take $\varepsilon = 1$.

By this construction of connection weights, the diagonal elements of $G$ are always normalized to one, thus preventing the coupling to be arbitrary large or small. Although $G$ becomes asymmetric for any value of $\alpha$, it can be written as $G = D_l W D_r$, where $W$ is a zero row-sum matrix with off-diagonal elements $W_{ij} = -\rho_{ij}$, $D_l = \text{diag}\left( 1/\sum_{j \in N_1} \rho_{1j}, \ldots, 1/\sum_{j \in N_N} \rho_{Nj} \right)$ and $D_r = \text{diag}\left( (\varepsilon + C_1^\alpha), \ldots, (\varepsilon + C_N^\alpha) \right)$. It can easily be shown that the eigenvalues of $G$, $\lambda_i$ ($i =$ 1,...,$N$), are the same as the eigenvalues of $D_r^{1/2}D_l^{1/2}WD_l^{1/2}D_r^{1/2}$, i.e. real and non-negative with smallest eigenvalue as $\lambda_1 = 0$. For this case, Gerschgorin circle theorem [25] guarantees that $0 < \lambda_2 \leq ... \leq \lambda_N \leq 2$. Another important term concerning our procedure is the effect of different values of $\alpha$. It is worth mentioning that the case with $\alpha = 0$ corresponds to the optimal situation proposed in [17], which itself has the optimal case of [18, 19] as a special case. For large absolute values of $\alpha$, it may happen that the resulting weighted network is approximately disconnected [17] and thus $\lambda_2$ is close to zero; therefore, we limit ourselves to $\alpha$ with small absolute values.

### III. NUMERICAL SIMULATIONS

#### A. Performance in artificially constructed networks

By sweeping $\alpha$ and calculating the values of $\lambda_N/\lambda_2$ we can study the synchronizability profile of dynamical networks with different topological properties such as scale-free, Watts-Strogatz and random networks. We will construct scale-free networks using an algorithm proposed in [17], which itself is a generalization of the preferential attachment growing procedure introduced in [7]. Namely, starting with a network of $m + 1$ all-to-all connected nodes, at each step a new node is added with $m$ links that are connected to node $i$ with probability $p_i = (k_i + B)/\sum_j(k_j + B)$, where $k_i$ is the degree of the node and $B$ a tunable real parameter controlling the heterogeneity of the network [17]. Watts-Strogatz networks with average degree $<k> = 2m$ are constructed based on the Watts-Strogatz algorithm [6] with probability of rewiring $P$. Since we are also interested in studying networks with small mean degree (e.g. $<k> = 2$), and construction of connected Watts-Strogatz networks with $m = 1$ for large $N$ is difficult, we study the $\lambda_N/\lambda_2$ for the largest connected component of such networks. We also consider a class of connected random networks with predefined average degree, where in order to build the network with $N$ nodes and exactly $mN$ edges, i.e. $<k> = 2m$, first $[mN/2]$ of possible $N(N–1)/2$ edges are selected randomly, which results in $Q$ connected components. Then, these connected components are randomly connected through ($Q$–1) edges (if ($Q$–1) > [$mN/2$], the network is rejected). Other remaining edges are selected randomly, which results in a connected random graph with exactly $<k> = 2m$.

Fig. 1 shows the logarithm of the eigenratio $\lambda_N/\lambda_2$ in the parameter space ($\alpha,B$) for scale-free networks of different size and topological properties. For the case with $m = 1$ by increasing $\alpha$, $\lambda_N/\lambda_2$ is rapidly decreasing, i.e. synchronizability is enhanced. Note that $\alpha = 0$ recovers the optimal condition of [17], which itself has the optimal situation of [19] as a special case. For $m = 2$, the situation is somewhat different; for small values of $B$, $\lambda_N/\lambda_2$ is decreased by increasing $\alpha$. However, for larger values of $B$ (less heterogeneity in the degree distribution), there is a local minimum in $\alpha \sim 0.5$, and then by increasing $\alpha$, the eigenratio is also increased and by further increasing $\alpha$ over a value around 1, the eigenratio starts decreasing. Although $\alpha = 1$ is not the exact optimal point, to avoid the network from being disconnected (high values of $\alpha$ may lead the network to be disconnected), we consider $\alpha = 1$ for weighting the scale-free networks in order to enhance their synchronizability. Considering node betweenness centrality in addition to edge load makes the resulting weighted network more homogeneous and thus enhances its synchronizability.

FIG. 1.

For comparison, we have also applied our weighting procedure to Watts-Strogatz networks that exhibit more homogeneity in the network structure than scale-free networks. We consider Watts-Strogatz networks with $m = 1$ and $m = 2$. For the cases with $m = 1$, we consider the largest connected component of the network, where its average degree is about $2.23 \pm 0.1$. Fig. 2a (Fig. 2c) shows the logarithm of $\lambda_N/\lambda_2$ as a function of $\alpha$ and $P$ for largest connected component of Watts-Strogatz networks with $m = 1$ and $N = 1000$ ($N = 2000$). The eigenratio profile for Watts-Strogatz networks with $m = 2$ ($<k> = 4$) is shown in Fig. 2b and Fig. 2d for $N = 1000$ and $N = 2000$, respectively. As it is seen, there is a clear optimum in $\alpha \sim 1$. Also, the effect of the Watts-Strogatz phenomenon on the synchronizability of the network is clearly seen from these graphs, i.e. the synchronizability of the network is greatly enhanced by introducing some rewirings.

FIG. 2.

Here we also report the behavior of the synchronizability of the weighted networks with $\alpha = 1$ as a function of network size $N$. Fig. 3a shows the results for scale-free networks with different values of $B$ and random networks, all the cases with $m = 1$. As expected, the propensity of synchronization for weighted scale-free with lower values of $B$ is better regardless of $N$ [16]. Also, weighted scale-free networks exhibit better synchronizability than random networks. For $m = 2$ the profile of $\lambda_N/\lambda_2$ as a function of $N$ is depicted in Fig. 3b, where again weighted scale-free networks show better synchronizability than Watts-Strogatz and random networks. It also illustrates the fact that synchronizability of the weighted networks is almost independent of network size. Indeed, the average degree seems to be the only important factor affecting the synchronizability of the weighted networks.

FIG. 3.

## B. Performance in real-world networks

Although we showed that applying (2) for weighting the edges of dynamical networks greatly enhances the synchronizability of the network, i.e. reduces the eigenratio $\lambda_N/\lambda_2$, many real-world networks cannot be simply modeled by these scale-free, Watts-Strogatz or random network models. Real-world networks may possess a number of complex topological properties, which can make it difficult to construct a model that mimics all of them. Thus, we consider some available real-world undirected networks[1] and apply the proposed weighting algorithm (2) to study the behavior of $\lambda_N/\lambda_2$ in the resulting weighted networks. We consider some real-world networks including protein structure network [26], network of power grid [6], US airport network[2] [27], Email network [28], protein-protein interaction network [29], yeast protein interaction network [30], Internet on the level of autonomous system [31], and the network of coauthorship [32].

---

[1] All of these networks are downloadable from Internet in the web site provided by the authors of the original works; interested readers may refer to the cited work.

Table I summarizes the results. For all networks, our proposed weighting algorithm is the best. Let us remark that since for the algorithm of [20] there is no optimal value of $\beta$, thus for each network we adopt the least $\lambda_N/\lambda_2$ among the cases with $\beta = 1$, $\beta = 2$, and $\beta = 3$. For the methods of [18] and [17] we used the optimal condition $\beta = 1$ and for our proposed method we used $\alpha = 1$. Since the method of [18] is a simple scaling based on the degree of the nodes, it is incapable of capturing all the useful information, and thus, its performance is always worse than that of the one proposed by [17] that indeed considered the edge betweenness centrality as well as scaling based on the nodes. The method proposed in [20] uses the degree of nodes in a different way, and its performance is not always better than [17]. Indeed, it can be seen that by increase of the heterogeneity of the network, i.e. increase of the standard deviation of the node-degrees, the algorithm of [20] performs better than [17]. Our proposed algorithm is an intelligent extension of [17]; to consider the heterogeneity of the network, it considers the node betweenness centrality in addition to the edge betweenness centrality. Thus, it always outperforms to [17]. However, for networks with high levels of heterogeneity, [20] gives results better than [17] and closed to our results.

TABLE I.

---

[2] The original version of the US airport network is a weighted network, but here we have considered only the unweighted version.

## C. Enhancing phase synchronizability in coupled nonidentical dynamical systems

The rational behind taking the eigenratio $\lambda_N/\lambda_2$ as a synchronizability measure of dynamical networks is the master-stability-function [21]. Since the theory of the master-stability-function was developed for local stability of synchronization of identical dynamical systems, it can not be directly applied to coupled nonidentical systems. However, coupled nonidentical dynamical systems may exhibit some weaker types of synchronization such as phase synchronization [33]. Here, we study the collective behavior of nonidentical Rössler oscillators [34] on scale-free, Watts-Strogatz and random networks to study how much the proposed weighting algorithm can improve the degree of phase synchronization. The dynamics of motion is governed by (2), but with nonidentical individual dynamical systems, where the dynamics of the $i$-th node with state vector $\mathbf{x}_i = (a_i, b_i, c_i)$ is given by $\mathbf{F}_i(\mathbf{x}_i) = [-\omega_i b_i - c_i, \omega_i a_i + 0.165 b_i, 0.2 + c_i(a_i - 10)]$, and $\mathbf{H}(\mathbf{x}) = a$ [17]. Here $\omega_i$ is the natural frequency of the $i$-th system that is randomly chosen from a Gaussian distribution with mean value $\omega_{mean} = 1$ and standard deviation $\Delta\omega = 0.03$.

To study the phase synchronization among coupled oscillators one can monitor the order parameter [8, 33] $\Phi = \left\langle (1/N) \sum_{j=1}^{N} e^{i\varphi_j(t)} \right\rangle_t$, where $\varphi_j(t) = \arctan(b_j(t)/a_j(t))$ represents the instantaneous phase of the $j$-th oscillator, and $\langle \cdot \rangle_t$ makes time averaging. The values $\Phi \approx 1$ indicate that the systems are phase synchronized. Behavior of $\Phi$ as a function of the general coupling strength $\sigma$ is shown in Fig. 4a for $m = 1$ and Fig. 4b for $m = 2$ using different networks topologies with $N = 1500$. For all of the cases, the weighting approach (2) with $\alpha = 1$ (solid lines) results in better phase synchronizability than the case with $\alpha = 0$ (dashed lines). This improvement is well pronounced for the case with $m = 1$ (Fig.4a). These results confirm that the master-stability-function gives also valid information for nonidentical oscillators. Note that synchronizability is improved if the $\sigma$ range for which $\Phi \approx 1$ increases.

FIG. 4.

## IV. CONCLUSIONS

By considering the eigenratio $\lambda_N/\lambda_2$ of the Laplacian of the connection as the synchronizability measure, we proposed a procedure for assigning proper connection weights to enhance the synchronizability of dynamical networks. To form the weights, we used the information of the node and the edge betweenness centrality measures. The algorithm was tested on artificially constructed networks such as scale-free, Watts-Strogatz and random networks as well as on some real-world networks. This method was also powerful in enhancing the phase synchronizability in networks of nonidentical dynamical systems.

## ACKNOWLEDGMENTS

This work has been supported by Swiss National Science Foundation through grants No 200020-117975/1 and 200021-112081/1.

TABLE I. Enhancing the synchronizability of some real-world networks with different algorithms. First column: the name of the networks. Second, third, and fourths columns: network size $N$, average node-degree $<k>$, and standard deviation of node-degrees std($k$), respectively. Fifth to eighth columns: the eigenratio $\lambda_N/\lambda_2$ using different weighting algorithms.

| Real-world networks | $N$ | $<k>$ | std($k$) | $\lambda_N/\lambda_2$ [18] | $\lambda_N/\lambda_2$ [17] | $\lambda_N/\lambda_2$ [20] | $\lambda_N/\lambda_2$ using (2) |
|---|---|---|---|---|---|---|---|
| Protein structure network | 95 | 4.48 | 1.45 | 392.7 | 63.1 | 262.2 | 23.1 |
| Power grid network | 4941 | 2.67 | 1.79 | 7349.1 | 393.2 | 14924.1 | 157.2 |
| US airport network | 500 | 11.92 | 22.36 | 63.4 | 11.5 | 4.8 | 2.9 |
| Email network | 1133 | 9.62 | 9.34 | 14.6 | 8.6 | 5.8 | 5.4 |
| Protein-protein interaction network | 2840 | 2.92 | 8.73 | 86.5 | 34.9 | 41.6 | 16.5 |
| Yeast Protein interaction network | 1458 | 2.68 | 3.45 | 238.1 | 52.4 | 269.1 | 25.6 |
| Network of Internet as autonomous system | 8689 | 4.08 | 29.08 | 55.8 | 13.9 | 3.6 | 3.1 |
| Coauthorship network | 4380 | 3.25 | 3.55 | 383.7 | 68.3 | 273.1 | 38.7 |

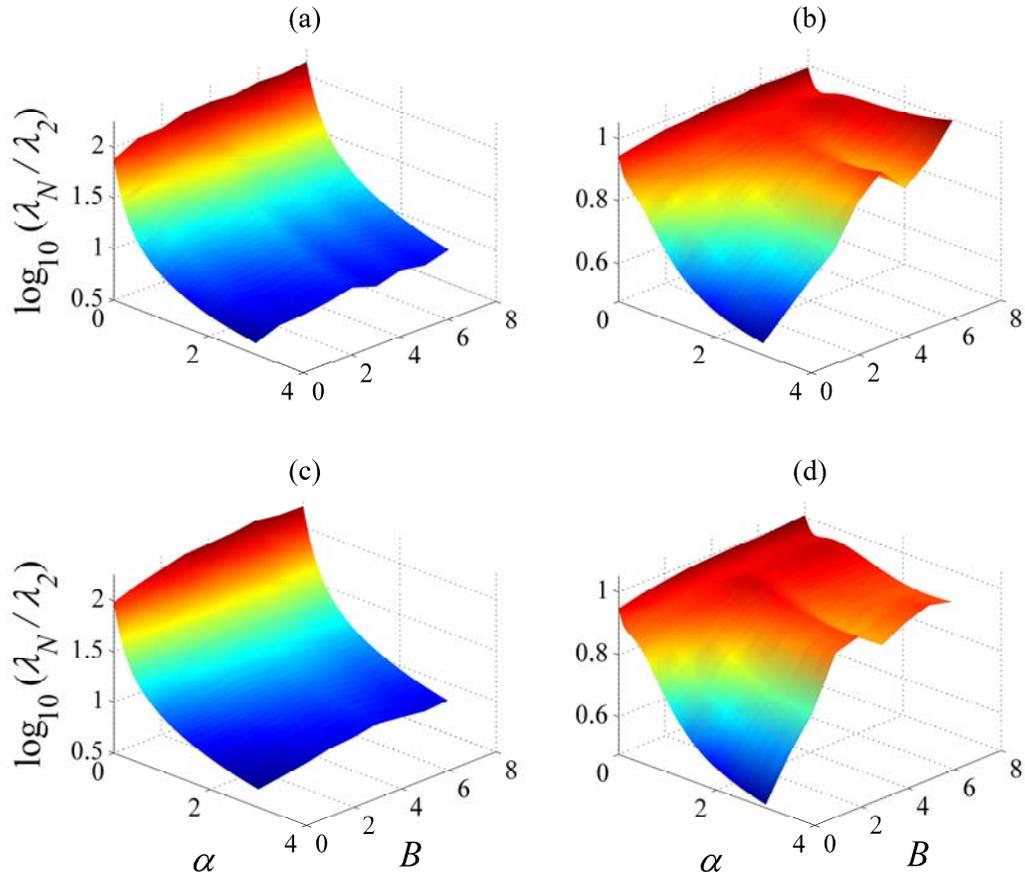

FIG. 1. (color online) Logarithm of the eigenratio $\lambda_N/\lambda_2$ as a function of ($\alpha$,$B$) for scale-free networks with a) $N = 1000$, $m = 1$, b) $N = 1000$, $m = 2$, c) $N = 2000$, $m = 1$, and d) $N = 2000$, $m = 2$. The graphs refer to averaging over 20 realizations of the networks.

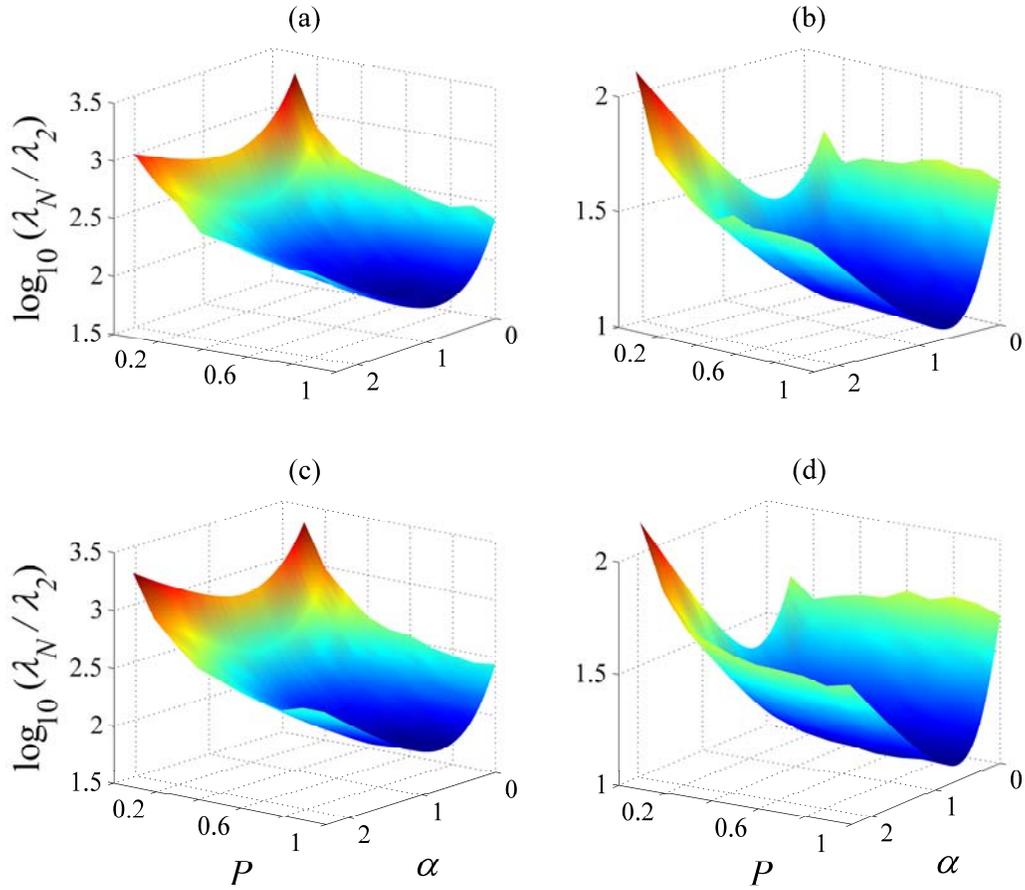

FIG. 2. (color online) Logarithm of the eigenratio $\lambda_N/\lambda_2$ as a function of ($\alpha$,$P$) for Watts-Strogatz networks with a) $N = 1000$, $m = 1$ (the largest component), b) $N = 1000$, $m = 2$, c) $N = 2000$, $m = 1$ (the largest component), and d) $N = 2000$, $m = 2$. The graphs refer to averaging over 20 realizations of the networks.

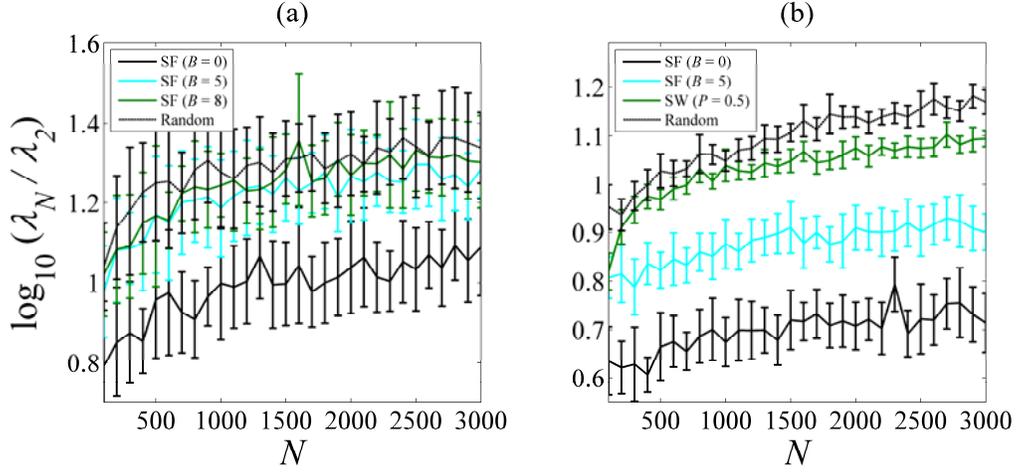

FIG. 3. (color online) Eigenratio $\lambda_N/\lambda_2$ (in logarithmic scale) as a function of the network size $N$ for different classes of networks (scale-free (SF), Watts-Strogatz (SW), and random) with $\alpha = 1$ and with a) $m = 1$, b) $m = 2$. Data refers to averages over 20 realizations.

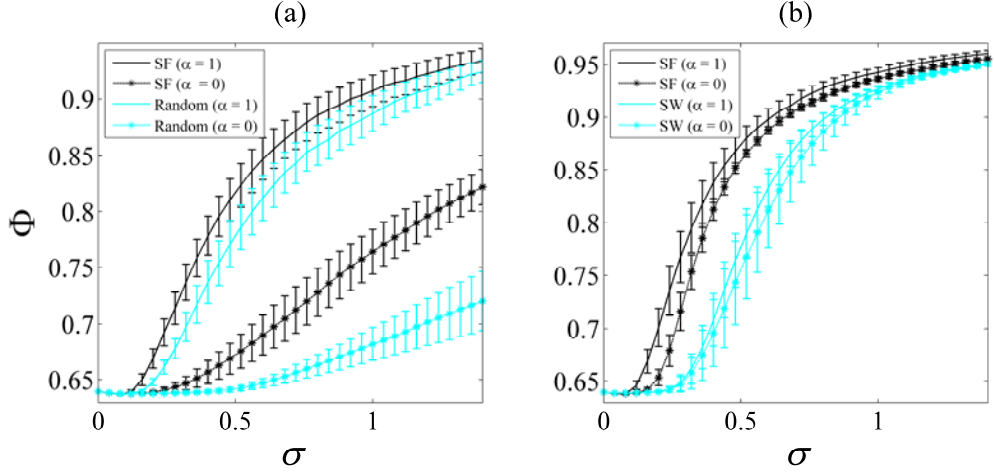

FIG. 4. (color online) The phase order parameter $\Phi$ as a function of uniform coupling strength $\sigma$ for coupled nonidentical chaotic Rössler oscillators. Data refers to averages over 20 realization of a) scale-free (SF) networks with $m = 1$ and $B = 5$, random networks with $m = 1$, b) scale-free networks with $m = 2$ and $B = 5$, Watts-Strogatz (SW) networks with $m = 2$ and $P = 0.5$. The network size in all case is $N = 1500$.